\documentclass[oldversion]{aa}
\usepackage{graphicx}
\begin{document}

\title{Discovery of a planetary nebula surrounding the symbiotic star DT Serpentis}

   \author{U.~Munari\inst{1},
	   R.~L.~M.~Corradi\inst{2,3},
           A.~Siviero\inst{4},
           L.~Baldinelli\inst{5},
           A.~Maitan\inst{5}
             }             

  \offprints{ulisse.munari@oapd.inaf.it}

  \institute{INAF Osservatorio Astronomico di Padova, 36012 Asiago (VI), Italy
  \and Instituto de Astrofisica de Canarias, E-38200 La Laguna, Tenerife, Spain
  \and Departamento de Astrof\'\i sica, Universidad de La Laguna, E-38206 La Laguna, Tenerife, Spain 
  \and Department of Physics and Astronomy, University of Padova, 36012 Asiago (VI), Italy 
  \and ANS Collaboration, c/o Astronomical Observatory, 36012 Asiago (VI), Italy}

   \date{Received YYY ZZ, XXXX; accepted YYY ZZ, XXXX}

  \abstract{We report the discovery of a planetary nebula centered on the
    poorly studied symbiotic binary star DT Ser.  In a few other symbiotic
    stars spatially resolved nebulae have been discovered as well, but
    only one of them is probably a genuine planetary nebula, while the
    others are likely to originate in complex mass-ejection episodes from
    the interacting binary central stars, possibly related to nova-like
    outbursts.  The rim of the planetary nebula around DT Ser is severely
    distorted toward a brighter star, 5 arcsec away.  In infrared WISE data,
    this star shows the presence of a detached cold-dust shell similar to
    those observed in post-AGB stars.  The apparent association of the
    symbiotic star and its planetary nebula with the nearby possible
    post-AGB object is discussed.  We also discuss the sparse and
    conflicting literature data that could support an observed variability
    of the surface brightness of the planetary nebula.  The puzzling and
    intriguing characteristics displayed by DT Ser are surely worth
    additional and more detailed investigations.

               }

   \authorrunning{U.Munari et al.}
   \titlerunning{Discovery of a planetary nebula around DT Ser}

   \maketitle

\section{Introduction}

Symbiotic stars are binaries composed of a late-type giant/supergiant and a
white dwarf (WD) companion.  The WD is bright (10$^3$ L$_\odot$) and hot
(10$^5$ K) and typically experiences surface stable H-burning of material
accreted from the cool giant.  The hard-radiation field ionizes part of the
wind of the giant, giving rise to a rich and high ionization emission line
spectrum and a bright Balmer continuum in emission, which are superimposed
on the absorption spectrum of the cool giant.  Given the usually large
distances to symbiotic stars and the limited extent of the region of the
cool giant wind ionized by the WD radiation field, nearly all known
symbiotic stars appear as point-like when viewed with a telescope.

Spatially resolved nebulae have so far been discovered only around a few of
the about 220 symbiotic stars listed in the catalog by Belczy{\'n}ski et al. 
(2000).  The 3D morpho-kinematic structure of some of them suggests a
ballistic launch by some sort of explosive event (Corradi 2003).  The
largest and more spectacular of these nebulae are seen in symbiotic binaries
where the cool giant is a Mira variable: R~Aqr (Lampland 1922, Solf and
Ulrich 1985, Paresce and Hack 1994), Hen 2-104 (Schwarz et al.  1989,
Corradi et al.  2001a, Santander-Garc{\'{\i}}a et al.  2008), BI Cru
(Schwarz and Corradi 1992), Hen 2-147 (Munari and Patat 1993,
Santander-Garc{\'{\i}}a et al.  2007), V1016 Cyg (Solf 1983, Corradi et al. 
1999) and HM Sge (Solf 1984, Corradi et al.  1999).  Nebulae have also been
discovered around symbiotic stars that harbor non-Mira cool giants: AS 201
(Schwarz 1991) and CH Cyg (Corradi et al.  2001b).  Bipolar planetary
nebulae like M 2-9 (Corradi et al.  2011) and Mz 3 (Guerrero et al.  2004;
Santander-Garc{\'{\i}}a et al.  2004) that may possibly host a symbiotic
nucleus (Schmeja and Kimeswenger 2001) are relevant bridges between
symbiotic binaries and aspherical planetary nebulae.

The discovery of extended nebulae around symbiotic stars, for its rarity, is
an interesting occurrence worth more detailed investigation.  The presence
and nature of such nebulae are relevant in addressing the age and formation
paths for symbiotic stars, their relation to planetary nebulae with binary
nuclei, and the feasibility of the symbiotic star channel for the
progenitors of type Ia supernovae.  In this paper we report our discovery of
a low surface-brightness planetary nebula, 11.4 arcsec in diameter, centered
on the symbiotic star DT Ser.

  \begin{figure*}
     \centering
     \includegraphics[width=18cm]{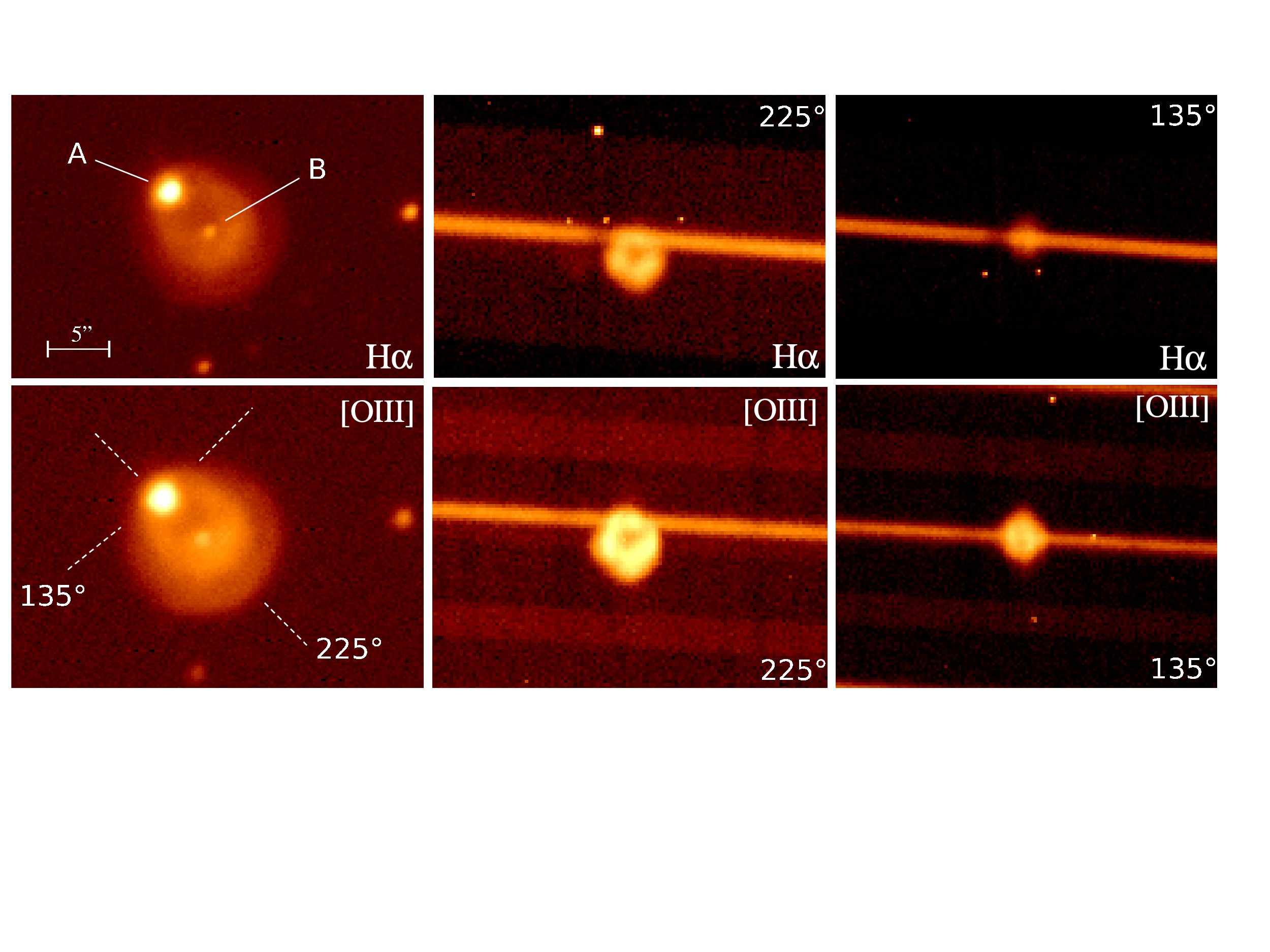}
     \caption{{\em Left}: NOT images (in H$\alpha$ and [\ion{O}{iii}] 5007 \AA\
     light) of the planetary nebula we discovered around the symbiotic star
     DT Ser.  North is up, east to the left, and the overall diameter of the
     nebula is 11.4 arcsec.  Stars A and B (5 arcsec away) that are mentioned in
     the text are identified.  The 225$^\circ$ and 135$^\circ$ slit
     positions, centered on star A, are also identified.  Positions S-E and
     N-W are referred to in Figure~4 and the text (sect.~5).  {\em Center}:
     portion of the original 2D Echelle spectra for the 225$^\circ$ slit
     position centered on H$\alpha$ and [\ion{O}{iii}] 5007 \AA.  {\em Right}: the
     same for the 135$^\circ$ slit position.}
     \label{fig1}
  \end{figure*}

\section{DT Serpentis}

DT Ser is a poorly studied variable, reported by the General Catalog of
Variable Stars (GCVS) to vary irregularly between magnitude 13.2 and 13.9 in
the blue photographic system, following early discovery reports of 1957. 
Jurdana-{\v S}epi{\'c} and Munari (2010) found DT Ser stable around mean
values of $B$=13.17 and $I_{\rm C}$=11.63 on Asiago archive plates exposed
between 1972 and 1977. USNO-B first- and second-epoch magnitudes of DT Ser
are $B$=14.11 and $B$=13.56, respectively.

Bond (1978, hereafter B78) reported in his spectroscopic survey of high galactic
latitude blue variables that DT Ser displayed a G-type absorption
spectrum and superimposed emission lines, in particular from [\ion{O}{iii}].  He
suggested a classification as a probable symbiotic binary, noted a stellar
appearance, and observed the presence to the south west of what he described
as a {\it slightly} fainter companion 5 arcsec away (which is merged with
the primary into an unresolved and elongated single stellar image on Palomar
Sky Survey plates).  In the following we call "A" the star
identified by B78 as the symbiotic star, and "B" the fainter companion 5
arcsec away (identified in Figure~1).

Cieslinski et al. (1997, hereafter C97) obtained $U$$B$$V$$R_{\rm C}$$I_{\rm
C}$ photometry and low-resolution spectroscopy of DT Ser.  In contrast to
B78, C97 found no emission lines associated to star A and instead determined
that the emission lines (\ion{He}{ii}, \ion{He}{i}, [\ion{O}{iii}]) were
associated with star B, which they identified as the symbiotic star.  C97
reported photoelectric photometry of star B as $V$=15.40, $U$$-$$B$=$-$0.59,
$B$$-$$V$=+0.51, $V$$-$$R_{\rm C}$=$-$0.06, and $R_{\rm C}$$-$$I_{\rm
C}$=$+$0.10, $\Delta V$=2.6 mag fainter than star A which C97 measured at
$V$=12.8 and $B$$-$$V$=+0.76.  PSF-fitting photometry on CCD images by
Henden and Munari (2001, 2008, hereafter HM01 and HM08 respectively)
measured star B at $V$=16.213, $U$$-$$B$=$-$0.820, $B$$-$$V$=+0.371,
$V$$-$$R_{\rm C}$=$-$0.055, and $R_{\rm C}$$-$$I_{\rm C}$=$+$0.11, and star
A at $V$=12.772, $U$$-$$B$=$+$0.109, $B$$-$$V$=+0.778, $V$$-$$R_{\rm
C}$=$+$0.462, and $R_{\rm C}$$-$$I_{\rm C}$=$+$0.475.  The difference in
magnitude between stars A and B as measured by HM08 is $\Delta V$=3.4, about
0.8 mag higher than derived by C97.

Neither B78 or C97 noticed any spatially resolved emission line on their
spectra, in spite of the long-slit instrumental set-up and the good seeing
conditions they must have enjoyed to be able to observe star B separately
from star A.  In sharp contrast, we first noticed the planetary nebula
around DT Ser directly at the telescope while cursorily inspecting the raw 2D
files of the spectra discussed in this paper: emission lines of uniform
spatial intensity were extending to one side of the stellar spectrum for
several times the seeing.  At the time of our observations, no trace of the
B star was noted on the TV guiding system of the telescopes, indicating that
it was several magnitudes fainter than star A.

\section{Imaging}

After the spectroscopic discovery of the spatially extended emission lines,
narrow-band images of DT Ser were obtained at the 2.6m Nordic Optical
Telescope (NOT) with the ALFOSC instrument on 30 March 2013.  The filter
central wavelengths and full width at half maximum (FWHM), and the
corresponding selected nebular emission lines, were 5007/30 \AA\
([\ion{O}{iii}] 5007) and 6563/30 \AA\ (H$\alpha$ with some contamination of
the [\ion{N}{ii}] 6548,6583 doublet, which is very weak in in DT Ser). 
Exposure time was 5 min in each filter.  The spatial scale of ALFOSC is
$0\farcs19$~pix$^{-1}$ and the seeing was $0\farcs7$ FWHM for the H$\alpha$
images and $1\farcs0$ for [\ion{O}{iii}].

The narrow-band images of DT Ser in Fig.~1 reveal the details of the
extended nebula, which has a diameter of 11.4 arcsec.  Its morphology is
very similar in the H$\alpha$ and [\ion{O}{iii}] filters and consists of an
inner, brighter rim and an outer, fainter attached shell.  This is a typical
configuration of planetary nebulae, in which the formation of the inner rim
is ascribed to the action of a fast post-AGB wind impinging on the slower,
denser AGB wind.  The wind interaction creates a hot inner bubble that
compresses gas and drives the expansion of the rim.  The attached shell,
which generally has a linearly decreasing surface brightness profile (as in
the case of DT Ser), is instead the result of the passage of the
photo-ionization front produced by the central star through undisturbed AGB
wind (see e.g.  Sch\"onberner et al.  2005).

  \begin{table}
     \centering
     \caption{Our photometry of DT Ser (merged images of stars A and B) and
     the corresponding total error budget (the table is published in its
     entirety in the electronic edition of the journal.  A portion is shown
     here for guidance regarding its form and content).}
     \includegraphics[width=8.5cm]{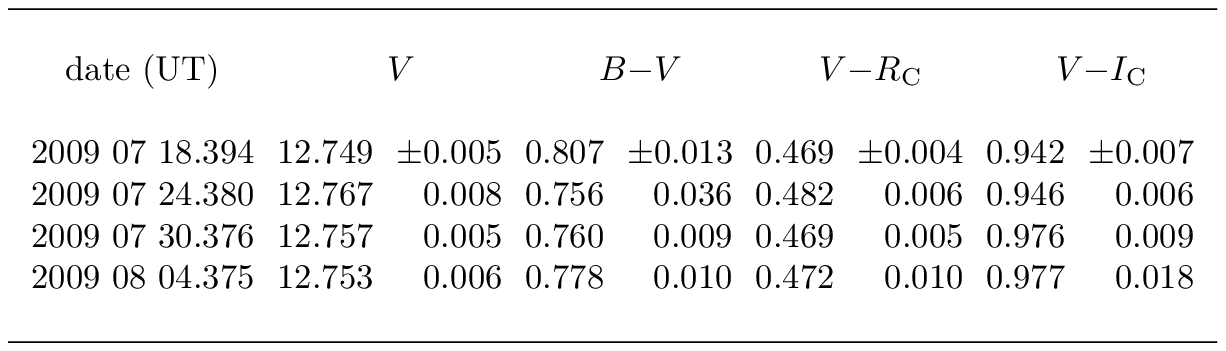}
     \label{tab1}
  \end{table}

  \begin{figure}
     \centering
     \includegraphics[width=8.9cm]{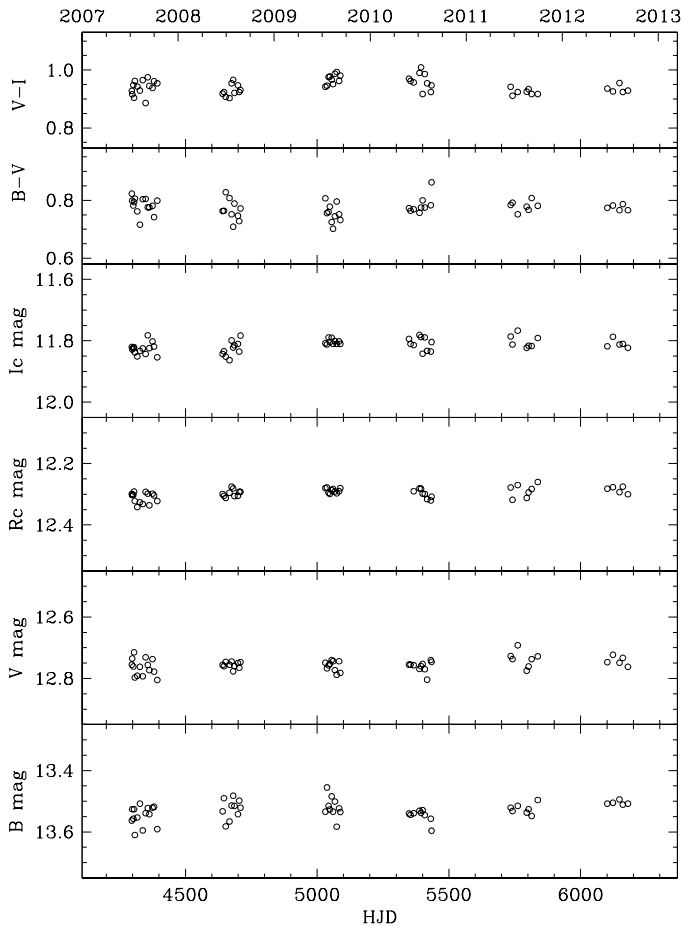}
     \caption{Photometric evolution of DT Ser (data from Table~1).}
     \label{fig2}
  \end{figure}
 
But the peculiar aspect of the nebula around DT Ser is that the rim
is severely distorted.  Its SE side is brighter and closer to the central
star, while the NE part is fainter and elongated toward star A, giving the
impression that it might be related to it.

\section{Photometric monitoring}

Since 2007, we have regularly observed DT Ser with the ANS Collaboration
telescope n.  020 as part of the monitoring of known symbiotic stars
described by Munari et al.  (2012a) and Munari and Moretti (2012).  The
telescope is a 0.40-m $f/5$ Newton reflector equipped with a HiSis~23ME CCD,
768$\times$1157 array, operated in binned mode for a scale on the sky of
1.24$\arcsec~{\rm pix}^{-1}$, and $B$$V$$R_{\rm C}$$I_{\rm C}$ filters from
Schuler.  With this telescope and typical seeing conditions, stars A and B
are merged into an unresolved single stellar image.  Instrumental
set-up and parameters of the photometric reduction (radius aperture, sky
annulus well external to the nebula) have been kept strictly stable during
the observing campaign, to avoid introducing spurious effects.  Our aperture
photometry was calibrated against the local photometric sequence of HM01.
It is given in Table~1 and refers to the integrated light of stars A and B.

  \begin{figure*}
     \centering
     \includegraphics[angle=270,width=16cm]{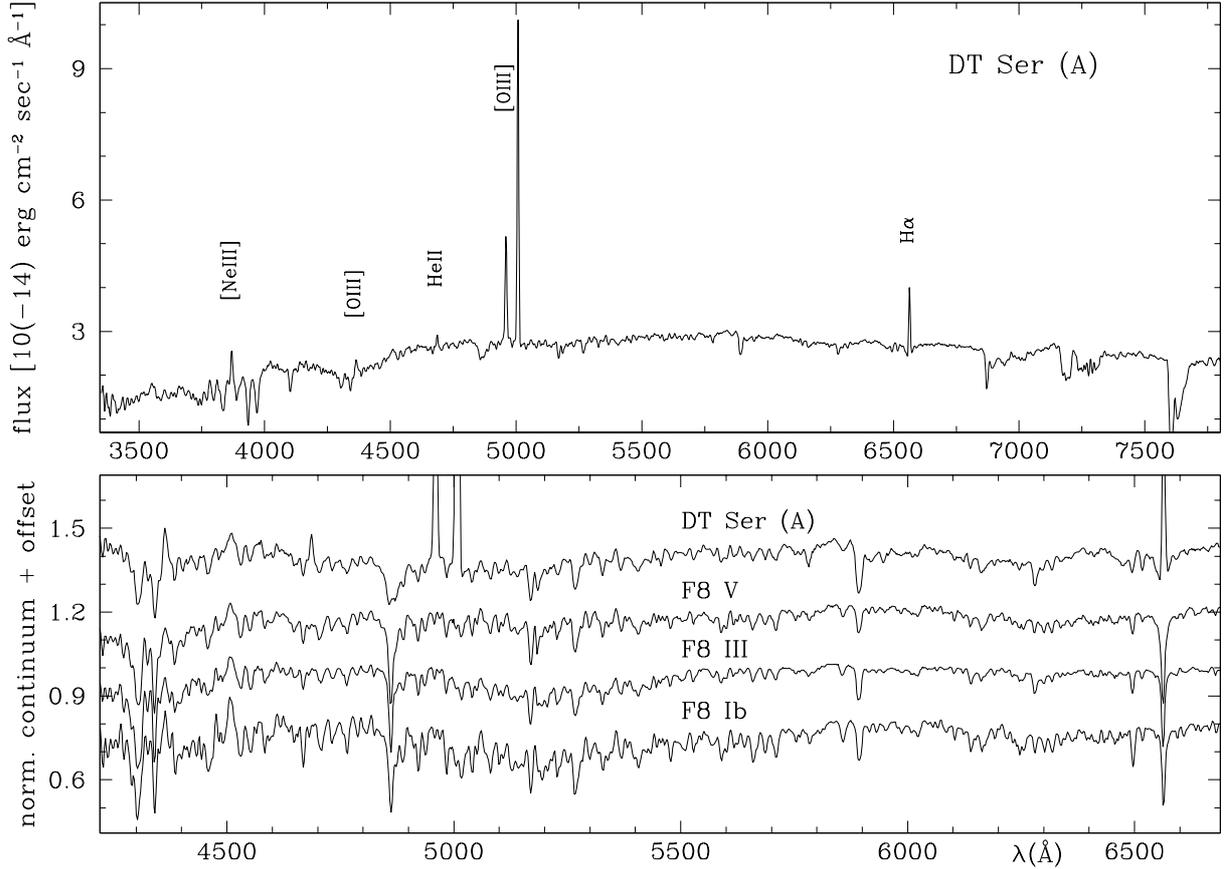}
     \caption{{\em Top}: Flux-calibrated spectrum of DT Ser (star A)
     for 25 Aug 2011.  {\em Bottom}: Comparison of the spectral continuum of
     DT Ser from the upper panel with those of MKK standards HD~9826 (F8~V),
     HD~224342 (F8~III), and HD~45412 (F8~Ib).}
     \label{fig3}
  \end{figure*}

  \begin{figure*}
     \centering
     \includegraphics[angle=270,width=16cm]{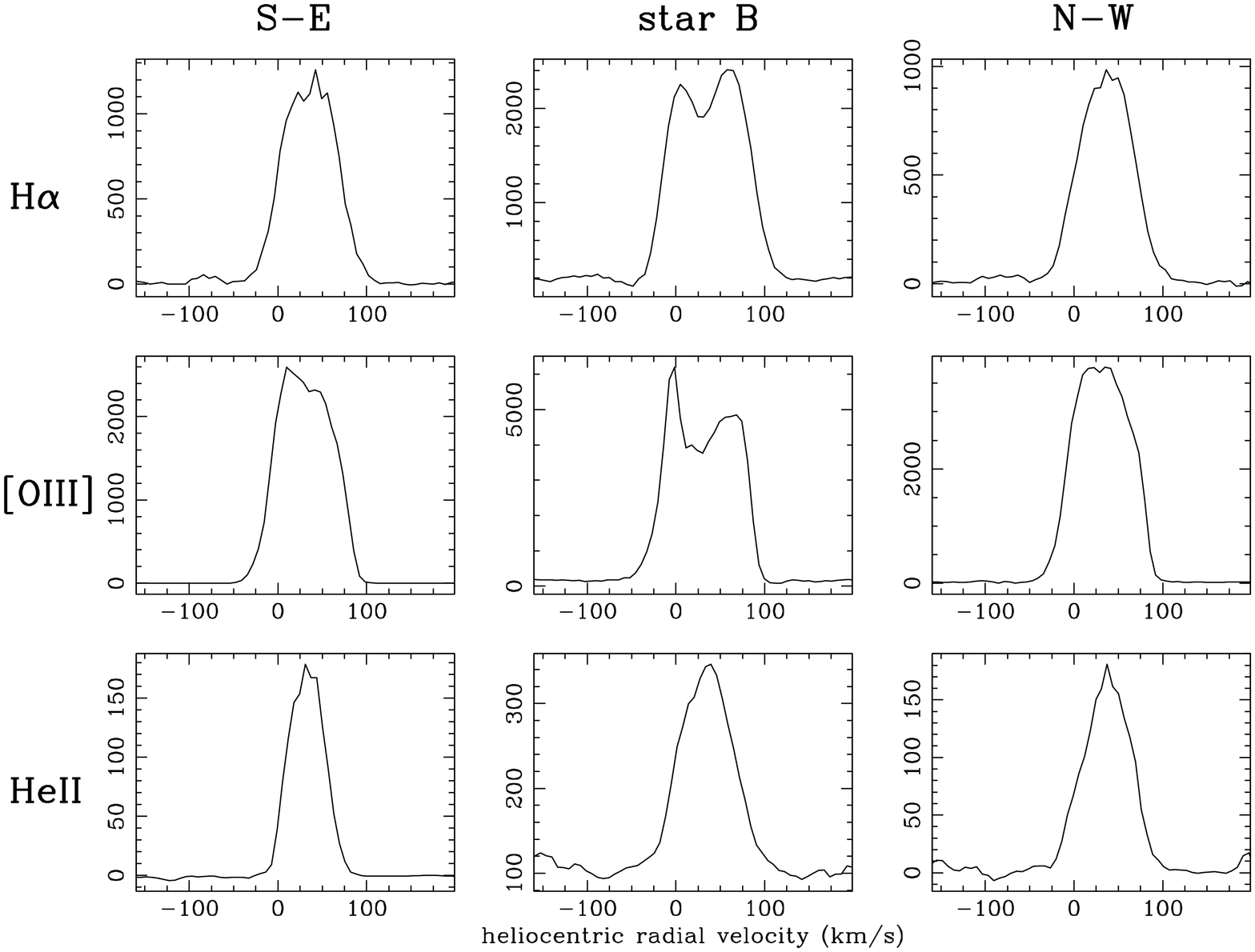}
     \caption{Velocity profiles of the planetary nebula emission lines
     from spatially resolved Echelle spectroscopy (see sect. 5). The 
     S-E, N-W, and star B positions are identified on the planetary nebula
     image at the top-left corner of Figure~1.}
     \label{fig4}
  \end{figure*}

The photometric evolution is presented in Figure~2, which indicates for all
bands a steady brightening with time of the blended image of stars A + B
over the 1883 day interval covered by the data in Table~1.  The brightening
amounts to $\Delta B$=$-$0.026, $\Delta V$=$-$0.021, $\Delta R_{\rm
C}$=$-$0.024 and $\Delta I_{\rm C}$=$-$0.026 mag.  Assuming that star A
remained constant at the level listed by HM01, star B
should have brightened accordingly from 18.88 to 17.29 in $B$, 18.10 to
16.69 in $V$, 17.71 to 16.31 in $R_{\rm C}$ and 16.92 to 15.27 in $I_{\rm
C}$.

  \begin{table}
     \centering
     \caption{Integrated fluxes (observed and de-reddened by $E_{B-V}$=0.20)
     of the emission lines in the low-resolution spectrum of DT Ser
     presented in Figure~3.}
     \includegraphics[width=5.5cm]{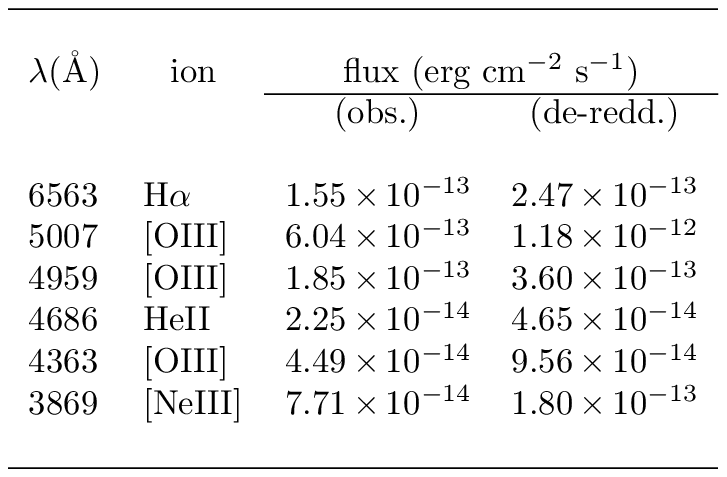}
     \label{tab2}
  \end{table}

\section{Spectroscopy and kinematics}

The low-resolution, flux-calibrated spectrum of DT Ser presented in Figure~3
was obtained on 25 Aug 2011 under 2.5-arcsec seeing conditions, with the
Asiago 1.22m telescope + B\&C spectrograph.  The 2-arcsec-wide and
5-arcmin-long slit was centered on star A and oriented E-W, thus
intersecting the distorted rim of the nebula.  The slit missed star B, which
was invisible on the telescope TV guiding system and, as expected, did not
reveal itself during the data reduction.  The sky background was traced from
12 to 22 arcsec away from star A in both direction, therefore well external
to the planetary nebula and star B.  The integrated fluxes of the emission
lines are listed in Table~2.  We have $\chi^2$ compared the continuum
spectrum of star A with the complete MK spectral atlas obtained with the
same Asiago 1.22m telescope, spectrograph, and detector.  When we masked the
regions affected by the emission lines, the best match was found around
spectral type $\sim$F8.  The fit is not satisfactory however, especially for
the luminosity class that remains essentially undetermined, and seems to
suggest that the absorption continuum of star A is not that of a {\em
normal} star.  The comparison of the absorption continuum of star A with
those of MKK standards for types F8~V, F8~III, and F8~Ib is presented in
Figure~3.

Spatially resolved, high-resolution Echelle spectroscopy of star A and the
surrounding nebula has been obtained on 27 Aug 2012 with the Asiago 1.82m
telescope, under 2 arcsec seeing conditions.  The spectrograph slit, widened
to 2 arcsec and providing a resolving power of 20\,000, extended spatially
for 30 arcsec.  Star A was placed at the center of the slit and the
telescope auto-guide tracked it.  Four separate deep spectra were obtained
with the slit rotated by $0^\circ$, $45^\circ$, $90^\circ$, and $135^\circ$
from the north-south alignment.  Small portions of these spectra centered on
[\ion{O}{iii}] 5007~\AA\ and H$\alpha$ are presented in Figure~1.  They
reveal the kinematical properties of the extended nebulosity, and in
particular the hollow elliptical position-velocity structure typical of rims
in planetary nebulae.

On each of the Echelle exposures, calibrated spectra were extracted tracing
star A and two parallel positions separated by $\sim$4 and $\sim$8 arcsec on
either sides of star A.  A total of 20 spectra were obtained: four referring
to star A, eight distributed across the planetary nebula, and another eight
symmetrically placed on the other side of star A which were used to check
the other spectra.  The velocity profile of the H$\alpha$, [\ion{O}{iii}]
5007 \AA, and \ion{He}{ii} 4686 \AA\ emission lines at three representative
locations of the planetary nebula are presented in Figure~4.  These three
points (identified in Figure~1) are aligned along the diameter of the nebula
and avoid the bright star A.  The heliocentric values of the photocentric
radial velocity of the emission line profiles, averaged over the eight
positions covering the extension of the planetary nebula, are +37.7 $\pm$0.5
${\rm km\,s}^{-1}$ for H$\alpha$, +29.6 $\pm$0.4 ${\rm km\,s}^{-1}$ for
[\ion{O}{iii}] 5007 and 4959 \AA, and +34.0 $\pm$1.0 for \ion{He}{ii}
(quoted uncertainties are the error of the mean).  These differences reflect
different formation regions for the different ions combined with the
aspherical nebular geometry.  The heliocentric radial velocity of the
nebula, averaged over the various emission lines, is +34 ${\rm km\,s}^{-1}$. 
The systemic velocity of the nebula calculated from the central wavelength
(mean value of the two peaks) of the H$\alpha$ split emission line where the
spectrum crosses the central star (star B) is similar, +33 ${\rm
km\,s}^{-1}$.  The heliocentric radial velocity of star A, derived from its
absorption spectrum is $-$26.7 $\pm$0.8 ${\rm km\,s}^{-1}$.

The emission line profiles observed at the center of the nebula (labeled
star B in Figure 4) suggest that the expansion velocity within the nebula
increases with distance from the central star.  The \ion{He}{ii} line,
forming closer to the central star, is the sharpest and displays a Gaussian
profile like that expected from a filled spherical gas distribution.  The
H$\alpha$ line, forming in a volume of space larger than the \ion{He}{ii}
region, is broader than the \ion{He}{ii} line and displays a double-peaked
profile with a velocity separation of 55 ${\rm km\,s}^{-1}$. 
[\ion{O}{iii}], forming preferentially in the lower density outer regions,
shows an even larger velocity separation between the peaks, about 75 ${\rm
km\,s}^{-1}$.

\section{Discussion}

\subsection{Reddening}

The reddening that affects star A can be derived from HM01
$U$$B$$V$$R_{\rm C}$$I_{\rm C}$ and 2MASS $J$$H$$K_s$ photometry, and the
intrinsic colors for F8 stars taken from Fitzgerald (1970), Bessell (1990),
and Straizys (1992).  The result is $E_{B-V}$=0.19 $\pm$0.02.  A similar
amount can be derived by simple geometry.  The reddening at the Galactic
poles is $E_{B-V}$$\sim$0.036 and the dust layer extends for $\sim$140 pc on
both sides of the the Galactic plane (averaging the many determinations
found in literature).  The corresponding total reddening integrated toward
star A is $E_{B-V}$=0.20 for a Galactic latitude $b$=+10.3 and a line of
sight that traverses the whole dust layer.

\subsection{Star A}

Assuming a reddening $E_{B-V}$=0.19 and a spectral type F8, the distance to
star~A is 400 pc for a star of luminosity class V, 1200 pc for class III,
and 22 kpc for class Ib, adopting the Hipparcos scale of absolute magnitudes
calibrated by Sowell et al.  (2007).  The 61 ${\rm km\,s}^{-1}$ difference
between the radial velocity of star~A and the planetary nebula surrounding
star~B indicates that the two are not related, unless the single-epoch
radial velocity we measured for star~A is perturbed by a large-amplitude
orbital motion around an unseen companion.

There are several puzzling features however that should not occur if
star A were a normal field star unrelated to star B and the nebula.

First, as noted in sect. 3, the rim of the planetary nebula is severely
distorted.  Its SE side is brighter and closer to the central star, while
the NE part is fainter and elongated toward star A, giving the impression
that it might be related to it.

Second, with normal field stars we do not usually experience such a poor
$\chi^2$ spectral fitting to primary MKK standard stars as we obtained for
star A, suggesting a {\em peculiar} nature and/or chemical partition.

Third, the WISE satellite medium-infrared all sky survey lists a bright
source (WISE J180152.26-012615.8) coincident with star A with magnitudes
W1=10.637 (3.35 $\mu$m), W2=10.616 (4.6 $\mu$m), W3=8.872 (11.6 $\mu$m), and
W4=4.071 (22.1 $\mu$m).  This large excess in the W3 and W4 bands indicates
the presence of a detached, cold-dust circumstellar shell around star A. 
The high spatial accuracy of the WISE satellite astrometry (0.06 arcsec on
both coordinates, which we confirmed by verifying the coincidence of WISE,
2MASS, and PPMXL coordinates for many field stars around DT Ser), rules out
the possibility that the WISE source could either be star B or the planetary
nebula.  Such a cold, detached dust shell is typically associated with
post-AGB objects.

If star A is indeed a post-AGB star, this could account for several
peculiarities we found.  ($a$) The $-$61 ${\rm km\,s}^{-1}$ difference in
radial velocity with the planetary nebula.  In post-AGB stars, the
pseudo-photosphere forms in the outflowing wind, which causes a blueshift of
the observed radial velocity.  An outflow velocity of about one hundred
${\rm km\,s}^{-1}$ is common in post-AGB stars.  ($b$) The difficulty in
classifying the spectrum of star A.  Absorption lines forming in a fast
outward-moving medium do not obey to the same curve-of-growth as those
forming in the nearly static atmospheres of normal stars.  The chemical
partition is also expected to be markedly different, because of the exposure
of nuclearly processed internal layers after the hydrogen-rich outer
envelope has been blown away.  ($c$) The discrepancy between the results of
Bond (1978), who saw Balmer lines of star A in emission, and those of
Cieslinski et al.  (1997) who instead found them in absorption. 
Emission lines are known to vary in post-AGB stars, depending
primarily on the intensity of the variable wind.

\subsection{The planetary nebula}

The emission lines from the high excitation ([\ion{O}{iii}]$\gg$H$\alpha$)
planetary nebula are very bright, and their spatial extension away from the
stellar image is quite obvious, as Figure~1 shows.  How could it be that
neither B78 nor C97 noticed them?  Especially puzzling is the non-detection
by C97 who observed under good seeing conditions and exposed {\em
separately} on stars A and B.  An unorthodox speculation would be to assume
that the surface brightness of the planetary nebula may have varied with
time.  This could happen in response to varying photo-ionization input from
the central star.  The timescale of the response from the nebula is the
recombination time, which is related to the electron density.

The critical density for nebular [\ion{O}{iii}] lines, which are highly
pronounced in DT Ser, is $\log N_e$=5.8 cm$^{-3}$, that for the missing
[\ion{N}{ii}] lines is $\log N_e$=4.9 cm$^{-3}$.  The [\ion{O}{iii}] ratio
(5007+4959)/4363=17.6 from Table~2 confirms that the nebula has a very high
density, namely $\log N_e$$\ge$5.64~cm$^{-3}$ assuming an electron
temperature $T_e$$\le$20000~K as typical of photoionized nebulae (Osterbrock
and Ferland 2006).  Such electron density is one or two orders of magnitude
higher than commonly found in planetary nebulae (Stanghellini \& Kaler
1989).  A much higher electron temperature, such as encountered in shock
fronts, would lower the electron density ($\log N_e$=4.0 cm$^{-3}$ for
$T_e$=40\,000~K).  The electron density estimate is relevant for the total
mass derived for the nebula.  Approximating the planetary nebula with a
sphere uniformly filled with hydrogen, its mass would scale with distance D
and filling factor $\xi$ as
\begin{equation} 
M_{\rm PN} = 0.55\, \xi\, \left(\frac{D}{\rm kpc}\right)^3 ~{\rm M}_\odot .
\end{equation}
The typical mass of planetary nebulae (Boffi \& Stanghellini 1994) is few a
tenths of a solar mass, while that of symbiotic stars is one or two orders of
magnitude smaller, with the notable exception of Hen~2-104 which is as
high as $\sim$0.1~M$_\odot$ (Santander-Garc{\'{\i}}a et al.  2008).  This
would suggest a distance to DT Ser not much larger than 1 kpc for $\xi$=0.4,
a value typical of planetary nebulae (Boffi \& Stanghellini 1994).

Assuming $\log N_e$=5.4 cm$^{-3}$, the recombination timescale for
hydrogen of planetary nebula in DT Ser is (Ferland 2003)
\begin{equation} 
t_{\rm rec} = 0.66 \left(\frac{T_{\rm e}}{10^4 {\rm ~K}}\right)^{0.8} 
\left(\frac{n_{\rm e}}{10^9 {\rm ~cm}^{-3}}\right)^{-1}
\approx 85 {\rm ~days} ,
\end{equation}
which is shorter than the time scale between the sparse
observations available of DT Ser, and therefore compatible with
possile significant variation of the surface brightness of the
planetary nebula in between them.

Some studies have found by combining radio and optical data (e.g.  Lee \&
Kwok 2005; V{\'a}zquez et al.  1999) that a non-uniform internal
distribution of dust can play an important role in the morphology of some
PN.  The amount of dust associated with the PN in DT~Ser seems negligible
because of two reasons.  The sensitive WISE mid-infrared survey did not
detected any emission associated to the PN, and no \ion{Na}{i} doublet
absorption line is observed at the systemic velocity of the PN on our
Echelle spectrum of star A.  This spectrum has a signal-to-noise of 92 on
the continuum around the \ion{Na}{i} doublet at 5890 and 5896 \AA, and the
upper limit of the equivalent width of any PN-related component is 0.006
\AA.  Adopting the calibration of Munari \& Zwitter (1997), which was
derived for the diffuse interstellar medium, this translates into un upper
limit of $E_{B-V}$$<$0.002.  If star A is related to the planetary nebula
and it is seen through the outer layer of the nebula, the $E_{B-V}$$<$0.002
limit suggests a negligible amount of dust associated to the PN and located
along the line of sight to star A.

\subsection{Star B}

The planetary nebula we have discovered shows widespread \ion{He}{ii}
emission, indicating a hot central ionizing source: to support production of
\ion{He}{ii} via photo-ionization, the WD must have a temperature $T_{\rm
eff}$$>$55\,000 K (Allen 1984).  The optical photometric colors of very hot
stars do not strongly depend on their actual temperature, their effective
wavelengths being on the Rayleigh-Jeans tail of the energy distribution: for
an unreddened very hot star we may assume, according to Straizys (1992) and
Drilling and Landolt (2000), $U$$-$$B$=$-$1.20, $B$$-$$V$=$-$0.35 and
$V$$-$$I_{\rm C}$=$-$0.34.  These colors cannot be simultaneously reconciled
with those observed for star B ($U$$-$$B$=$-$0.71, $B$$-$$V$=$+$0.44 and
$V$$-$$I_{\rm C}$=$+$0.05, average of C97 and HM08 data) for any assumed
reddening. A compromise could be to settle for $E_{B-V}$$\sim$0.52, which is
appreciably higher than the interstellar reddening derived in sect.  6.1
above.  This suggests that star B is a binary system in which the ionizing
source has a cooler companion that softens the combined optical energy
distribution toward cooler temperatures, as in symbiotic binaries.  If
the companion is a star with a strong wind, part of it should be ionized by
the hot star, producing emission lines.  No sharp emission line component is
observed to stand over the planetary nebula diffuse emission at the position
of star B in the high-resolution Echelle spectra of Figure~1.

The most puzzling aspect of star B is its variability, which is not expected
from the central star of a normal planetary nebula.  The C97 and HM08
photometry of star B, separated in time by four years, differs by 0.8 mag in
$V$.  It has to be remarked that C97 performed aperture photometry, while
HM08 carried out PSF fitting because of the close proximity of stars A and
B.  Some light leak from star A could have contaminated the measurement of
star B by Cieslinski et al.  (1997), but with the available information we
cannot estimate if this indeed occurred, and if so, to what extent.  B78
reported that during acquisition of his spectra, star B appeared {\em only
slightly} fainter than star A.  When we were exposing our spectra, star B
was so much fainter than A as to be undetectable on the TV guiding systems
of both the Asiago 1.22m and 1.82m telescopes.

Without doubts, the brightness of the {\em combined} star A+B image does
indeed change with time, as outlined in sections 2 and 4 above.  The steady
increase in brightness in Figure~2 continued for five years.  If this is
related to the changing orbital aspect, the orbital period should be
$\geq$10 years.  This would resemble the low-amplitude ($\Delta V$=0.14
mag), 16.8 yr orbital modulation of the yellow symbiotic star V471 Per,
which harbors a G5~III donor star (Munari et al.  2012b).

\section{Concluding remarks}

The whole picture surrounding DT Ser is both puzzling and intriguing, and
the object is far more interesting and exciting than a normal symbiotic
star.  If there really is a symbiotic binary in the system, its association
with a planetary nebula is the first such known case.  If not, the large
photometric variability observed for the nucleus of the planetary nebula is
nevertheless a rarity, if not a unique case.  On top of all this, the bright
star A may also be a post-AGB star.  At 5 kpc distance, the 5 arcsec angular
separation translates into a spatial separation of 0.1 pc, the typical
radius of a planetary nebula.

A firm conclusion about the true nature of DT Ser, the planetary nebula, and
the two stars A and B needs additional devoted observations.  In particular,
it seems relevant to obtain ($a$) multi-epoch photometry and spectroscopy of
the spatially resolved star B to confirm its variability and look for a
signature of a companion; ($b$) deep imaging at higher spatial resolution of
the planetary nebula, especially the part closest to star A; ($c$) an
atmospheric analysis of star A, to determine the distance (from a comparison
of the derived $T_{\rm eff}$, $\log g$, and [M/H] with stellar isochrones)
and the evolutionary status (from chemical abundances); ($d$) epoch radial
velocities of star A to examine whether if the radial velocity difference
with the planetary nebula may be due to orbital motion (around an unseen
companion in a close orbit) or absorption lines forming in a optically thick
wind such as that of a post-AGB star.

\begin{acknowledgements}
Based in part on observations made with the Nordic Optical Telescope,
operated jointly by Denmark, Finland, Iceland, Norway, and Sweden, on
the island of La Palma in the Spanish Observatorio del Roque de los
Muchachos of the Instituto de Astrof\'\i sica de Canarias.
RLMC acknowledges funding from the Spanish AYA2012-35330 grant.
\end{acknowledgements}

\newpage
  \setcounter{table}{0}
  \begin{table*}
     \centering
     \caption{in extenso}
     \includegraphics[width=18cm]{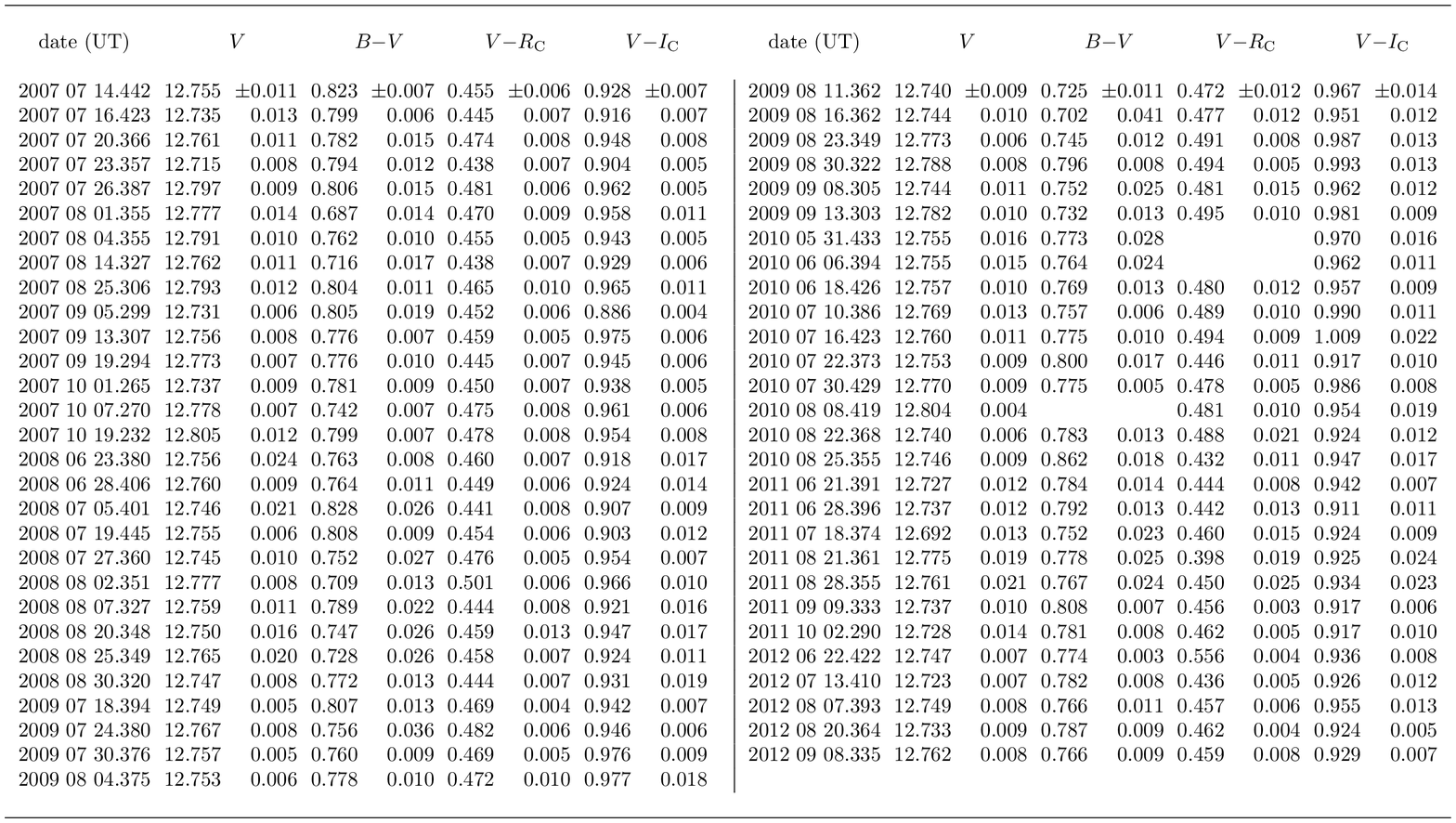}
  \end{table*}
\end{document}